\documentclass[reprint,prl]{revtex4-2}
\usepackage{xr-hyper}
\usepackage{tikz}
\usepackage[bookmarks=false]{hyperref} 
\usepackage{amsmath}
\usepackage{amssymb}
\usepackage{physics}
\usepackage{tensor}
\usepackage{graphicx}
\usepackage{bm}

\newcommand{\epsall}{\boldsymbol{\epsilon}}

\newcommand{\sigall}{\boldsymbol{\sigma}}

\def\onea{\tensor[^1]{A}{_1}}

\def\eps{\epsilon}

\def\onea{A_1^\circ}
\def\aone{A_1^\square}

\usepackage{xr-hyper}
\makeatletter
\newcommand*{\addFileDependency}[1]{
  \typeout{(#1)}
  \@addtofilelist{#1}
  \IfFileExists{#1}{}{\typeout{No file #1.}}
}
\makeatother

\newcommand*{\myexternaldocument}[1]{
    \externaldocument{#1}
    \addFileDependency{#1.tex}
    \addFileDependency{#1.aux}
}
\myexternaldocument{supplementary}

\begin{document}
\author{
    Mark A. Mathis$^1$, 
    Chris A. Marianetti$^1$
}

\address{$^1$ Department of Applied Physics and Applied Mathematics, Columbia University, New York, NY 10027}
\title{\textit{Ab initio} elasticity at finite temperature and stress in ferroelectrics}

\begin{abstract}
    Computing the temperature and stress dependence of the full elastic constant tensor from first-principles in non-cubic materials 
   remains a challenging problem. 
   Here we circumvent the aforementioned challenge via the generalized quasiharmonic approximation in conjunction with
   the irreducible derivative approach for computing strain dependent phonons using finite difference,
   explicitly including dipole-quadrupole contributions.
   We showcase this approach in ferroelectric PbTiO$_3$ using density functional theory,
   computing all independent elastic constants and piezoelectric strain coefficients at finite temperature and stress.
   There is good agreement between the quasiharmonic approximation and
   the experimental lattice parameters close to 0 K.
   However, the quasiharmonic approximation overestimates the temperature
    dependence of the lattice parameters and elastic constant tensor,
    demonstrating that a higher level of strain dependent anharmonic vibrational theory is needed. 
\end{abstract}

\maketitle

Ferroelectric materials have been widely studied due to technological
importance and interesting physics
\cite{Lines2001,Setter2006051606,Rabe2007,Cohen2008471,Liu2018041306}.
Many ferroelectric materials are band insulators, which are typically well described
by density functional theory (DFT).
For prototypical ferroelectric band insulators, the Born-Oppenheimer potential generated from
appropriate exchange-correlation approximations within DFT produce
temperature
dependent structural phase transitions consistent with experiment
\cite{Rabe19871987,King-Smith19945828,Zhong19956301,Rabe19962897,Waghmare19976161,Tinte2003064106,Nishimatsu2010134106,Wojdel2013305401,Paul2017054111,Masuki2022224104}, 
indicating an 
accurate representation of the vibrational free energy. 
However, evaluating piezoelectric properties at finite temperatures
and stress requires
the computation of relevant strain curvatures of the
DFT based vibrational free energy at finite temperatures and stress (i.e. the elastic constants). 
Computing these elastic constants requires encoding or sampling the vibrational
Hamiltonian as a function of strain, evaluating the vibrational free energy 
in some approximation
as a function of strain, and evaluating the second strain derivatives of the free energy.
Each of the aforementioned tasks presents substantial theoretical and computational challenges.

The standard approximation for computing finite temperature elastic constants
is the quasiharmonic approximation (QHA)
\cite{Togo2010174301,Wentzcovitch201099,Baroni201039}, 
yielding reasonable agreement with experimental measurements for a variety of
cubic systems 
\cite{Karki20008793,Ackland2003214104,Sha2006214111,Pham2011064101,Dragoni2015104105,Malica2020315902,Malica2021475901}. 
The computational cost of executing the QHA within DFT is appreciable \cite{Wu2011184115}, as 
evidenced by the sparsity of temperature dependent elastic constant
computations for non-cubic systems available in the literature
\cite{Shao2012083525,Bakare2022043803,Olsson2023111953}, 
and we are not aware of any published results at finite temperature and stress.
The aforementioned limitations can be mitigated by using the recently developed 
generalized quasiharmonic approximation \cite{Mathis2022014314}, 
which leverages the irreducible
derivative approach to computing phonons \cite{Fu2019014303,Bandi2023174302}.
Here we showcase the power of the generalized QHA by studying the displacive ferroelectric 
PbTiO$_3$ (spacegroup P4$mm$) using DFT, computing the 
lattice parameters, full elastic constant tensor, and piezoelectric strain
coefficients at finite temperature and
stress.
PbTiO$_3$ is an ideal candidate to study within the generalized QHA,
as the low symmetry ferroelectric phase persists to roughly $T=760$ K
\cite{Shirane1951265,Zhu2011,Jabarov20112300}.

DFT calculations were performed using the Vienna \textit{ab initio} simulation
package (VASP)
\cite{Kresse1993558,Kresse199414251,Kresse199615,Kresse199611169} with the
projector augmented wave (PAW) method \cite{Blochl199417953,Kresse19991758}
unless otherwise stated.
The generalized gradient approximation (GGA) revised for solids (PBEsol)
\cite{Perdew2008136406} and the strongly constrained and appropriately normed
(SCAN) \cite{Sun2015036402} exchange-correlation functionals were used. 
Convergence of the strain dependent phonons was achieved with a kinetic energy cutoff of 1000 eV and a
$\Gamma$-centered \textbf{k}-point mesh of 16$\times$16$\times$16 for
the primitive unit cell with corresponding mesh densities being used for
supercells. 
Details of the PAW potentials, finite difference calculations, and Fourier interpolation
are provided in Sec. I of the supplemental material (SM) \cite{SM}.
The computationally relaxed crystal structures using the PBEsol
and SCAN functionals are compared with low temperature experimental
measurements in Table \ref{table:latticeparams}. 
Due to the significant overestimation of the $c$ lattice parameter by the SCAN functional,
computations use the PBEsol functional unless otherwise stated.

\begin{table}
    \begin{tabular}{l lllll}
        \hline
        \hline
        Method & $a$ & $c$ & $z$(Ti) & $z$(O$_{1,2}$) & $z$(O$_3$) \\ 
        \hline
        \hline
        PBEsol         & 3.872 & 4.214 & 0.539 & 0.623 & 0.118 \\
        SCAN           & 3.865 & 4.341 & 0.545 & 0.638 & 0.139 \\
        \hline
        PBEsol         & 3.891 & 4.164 & 0.539 & 0.618 & 0.112 \\
        Mestric et al. & 3.891 & 4.168 & 0.542 & 0.629 & 0.124 \\
        \hline
        \hline
    \end{tabular}
\caption{
Lattice parameters and direct atomic coordinates along the $z$-direction.
(Top) Results of DFT relaxation using the PBEsol and SCAN functionals.
(Bottom) $T=1$ K QHA results compared with measurements at $T=12$ K \cite{Mestric2005134109}.
}
\label{table:latticeparams}
\end{table}

\begin{figure}
\begin{tikzpicture}
    \node at (0,9.75) { \resizebox{0.99\linewidth}{!}{\includegraphics{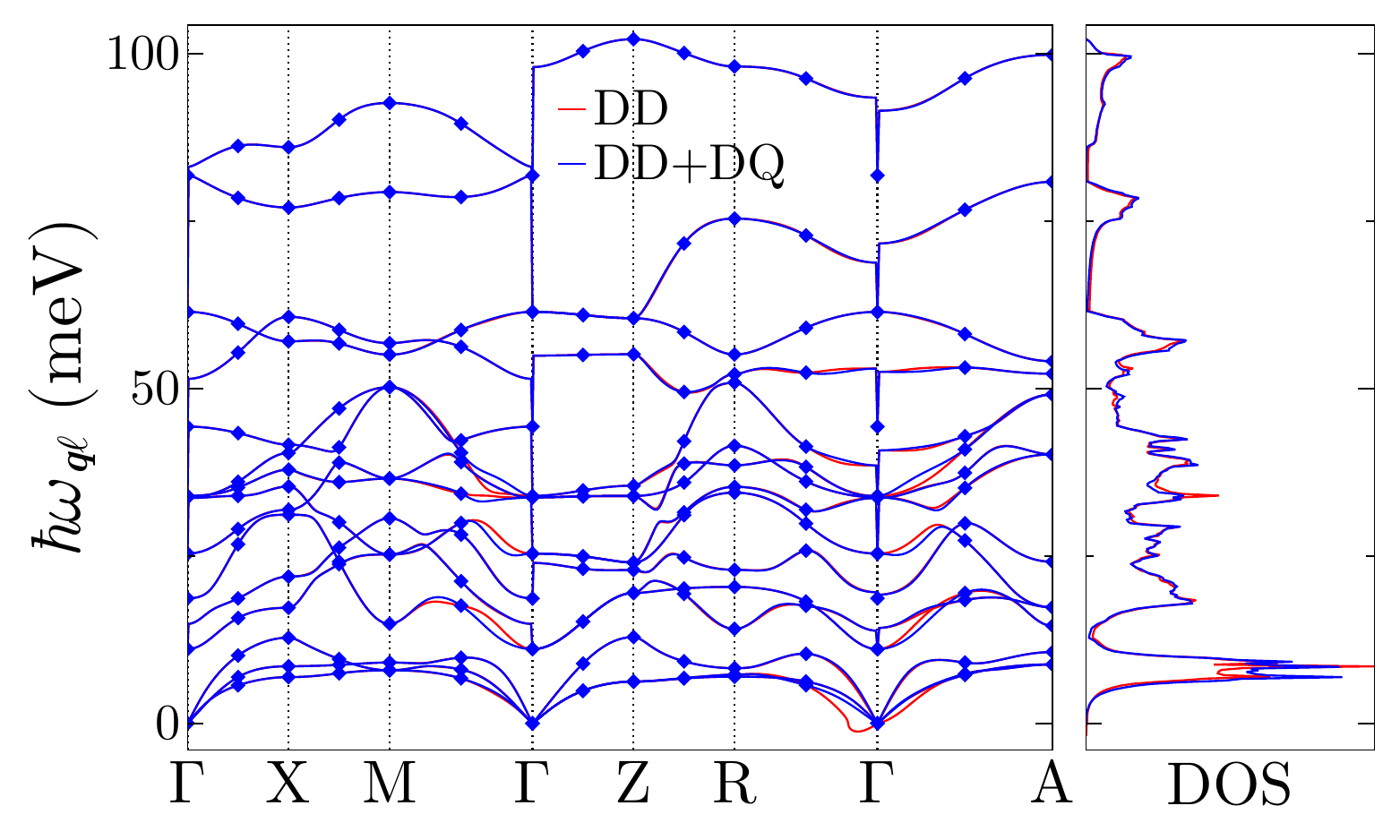}} };
    \node at (-3.9,11.95) {(a)};
    \node at (0.1,4.85) { \resizebox{0.965\linewidth}{!}{\includegraphics{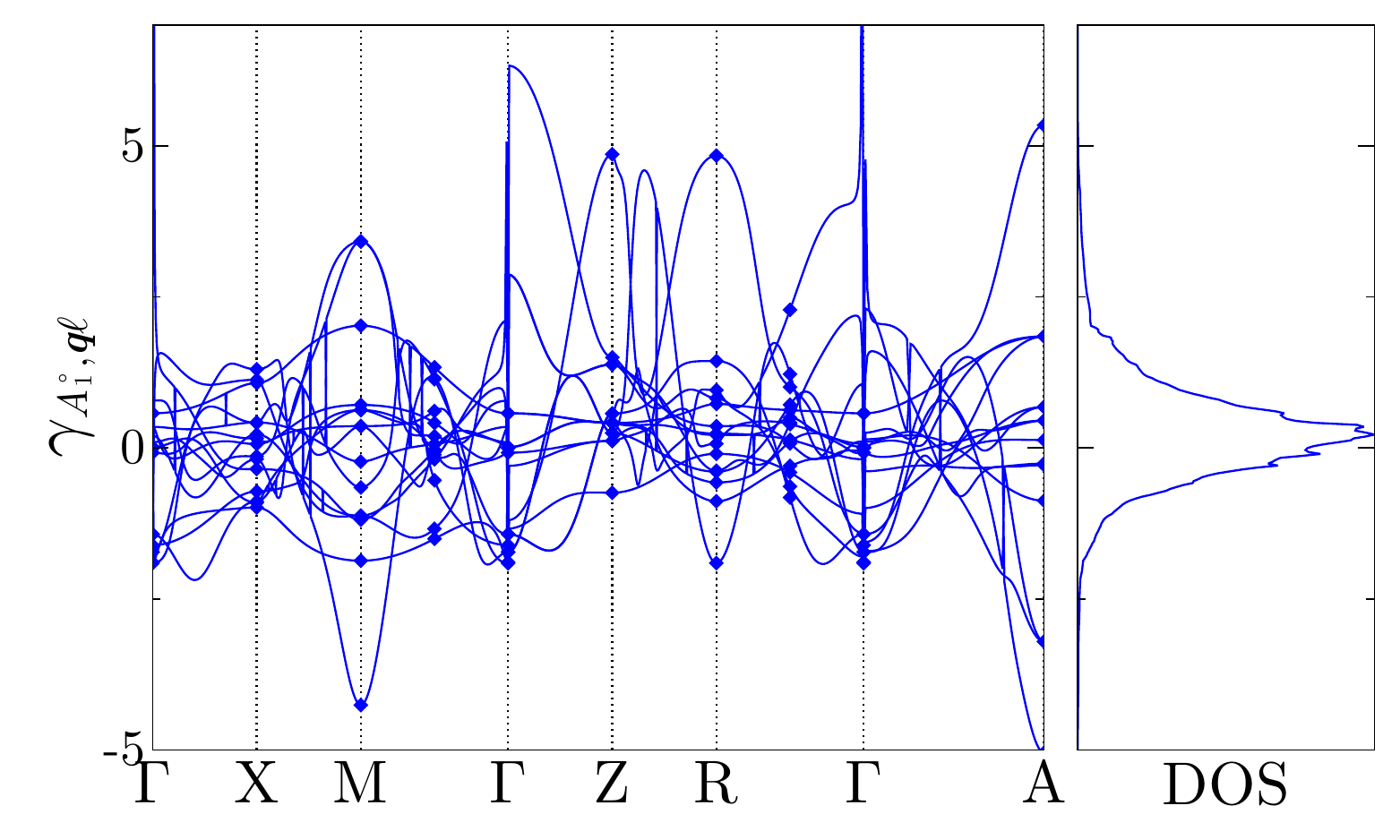}} };
    \node at (-3.9,7.05) {(b)};
    \node at (0,0) { \resizebox{0.98\linewidth}{!}{\includegraphics{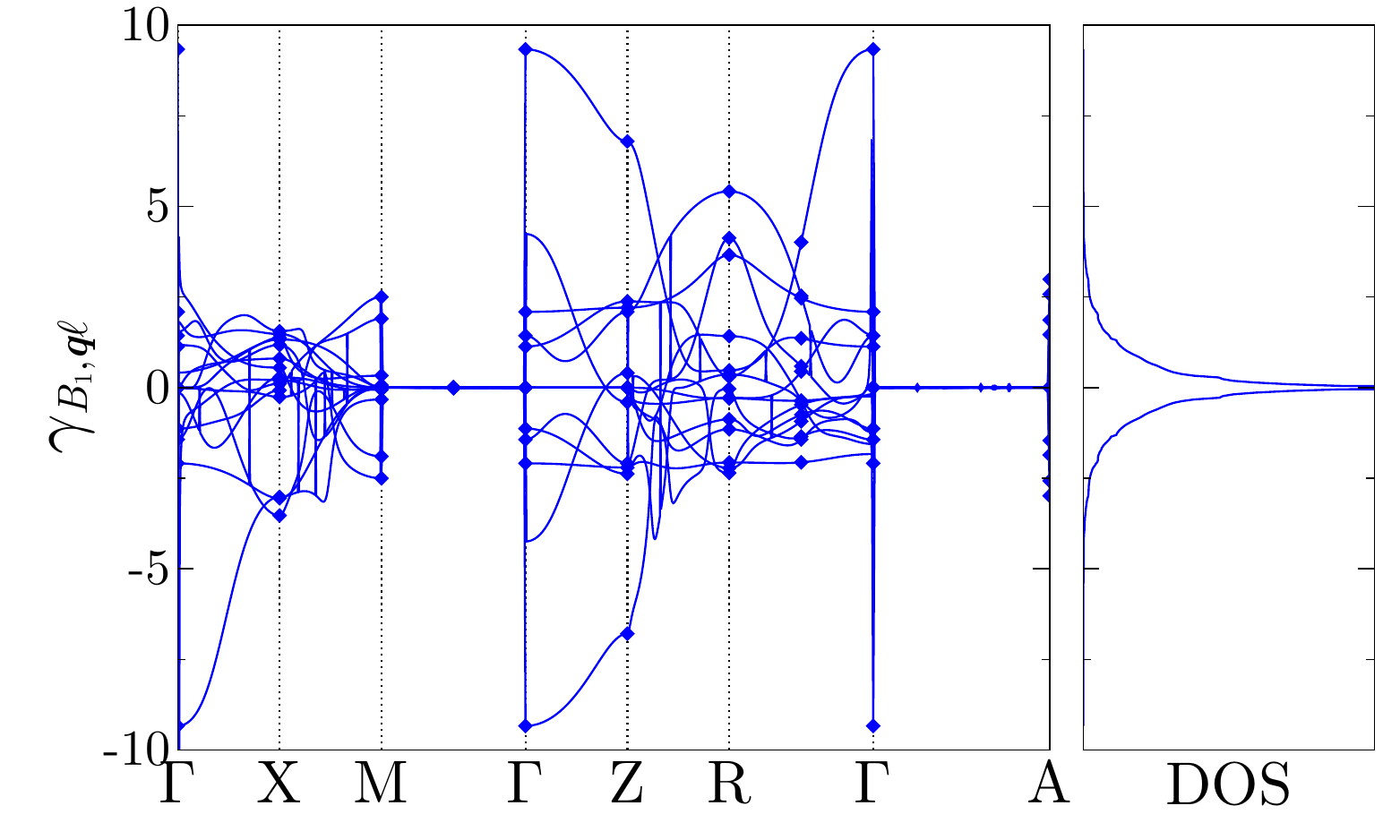}} };
    \node at (-3.9,2.3) {(c)};
\end{tikzpicture}
\caption{
    Phonons and specific Gruneisen parameters computed from DFT (diamonds) within the
    $4\hat{1}$ and $2\hat{S}_C$ supercells, respectively, and Fourier
    interpolated (lines). 
    Left panels show values along high symmetry directions, and
    right panels show the density of states.
    (a) Computed phonons which are Fourier interpolated with dipole-dipole (red) or dipole-dipole and
    dipole-quadrupole (blue) contributions.  
    Gruneisen parameters computed with strains transforming like the (b) $\onea$
    and (c) $B_1$ irreducible representations.
}
\label{fig:phonons}
\end{figure}
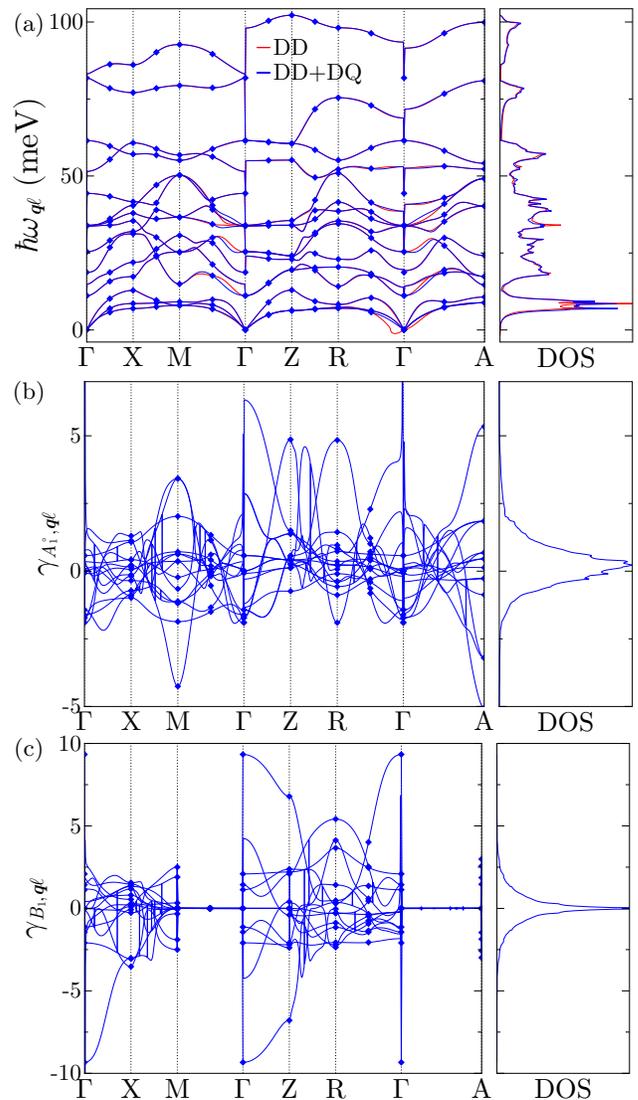

We begin by showcasing the computed phonons in the supercell
$\hat{S}_{BZ}=4\hat{1}$ (see Fig. \ref{fig:phonons} $(a)$), achieving good
agreement to previous computations \cite{Ritz2018255901}. 
The dipole-dipole contribution to Fourier interpolated phonons is included
\cite{Giannozzi19917231,Gonze199710355,Mathis2022014314} shown as the red lines,
where the dielectric tensor and Born effective charges were calculated from
density functional perturbation theory \cite{Baroni2001515,Gajdos2006045112}
within VASP.  
Along the path from the $\Gamma$ point to the R point, there are interpolated
imaginary phonon frequencies caused 
by a deficiency in the Fourier interpolation, which was elucidated in previous work \cite{Royo2020217602}. 
Supplementing the interpolation with the dipole-quadrupole interactions
has demonstrated the ability to remove these spurious imaginary frequencies from the 
interpolation \cite{Royo2020217602}.
Our Fourier interpolation of the phonons including the dipole-dipole and the
dipole-quadrupole contributions shown as the blue lines does not contain any
spurious imaginary frequencies, consistent with results in Ref. \cite{Royo2020217602}.
Computation of the dipole-quadrupole contribution has been implemented 
analogously to the dipole-dipole contribution, where 
dynamical quadrupoles were computed \cite{Royo2019021050} using density
functional theory implemented in the (ABINIT) package
\cite{Gonze2020107042,Romero2020124102} using the PBEsol optimized
norm-conserving Vanderbilt pseudopotential (ONCVPSP) \cite{Hamann2013085117}.
Our results illustrate that both dipole-dipole and dipole-quadrupole corrections
to the Fourier interpolation can straightforwardly be utilized in our irreducible derivative
approaches, which are based on the finite difference method.

We now present selected generalized Gruneisen parameters, 
\begin{align}
    \gamma_{i,\mathbf{q}\ell}=
    -\frac{1}{\omega_{\mathbf{q}\ell}}
    \frac{\partial \omega_{\mathbf{q}\ell}}{\partial \epsilon_i}, 
\end{align}
which encapsulate the first order strain dependence of the phonons along a
given strain (see Fig. \ref{fig:phonons} $(b)$ and $(c)$).
Computation of the full elastic constant tensor within the QHA requires the phonons 
to be computed as a function of all symmetrically distinct strains,
whereas thermal expansion computations only use strains associated with changes
in volume.
Thus, panel $b$ shows the $\onea$ Gruneisen parameter associated with expansion
along the $z$-axis, where the symmetrized representations of the identity strains
are denoted as
$\eps_{\aone}=\frac{1}{\sqrt{2}}(\eps_{xx}+\eps_{yy})$ and
$\eps_{\onea}=\eps_{zz}$.
Integration of the density of states yields averaged Gruneisen parameters of
$\bar{\gamma}_{\aone}$=1.78 and $\bar{\gamma}_{\onea}$=0.28.
Panel $c$ shows the $B_1$ Gruneisen parameter, where
$\eps_{B_1}=\frac{1}{\sqrt{2}}(\eps_{xx}-\eps_{yy})$.
Symmetry selection rules and first order perturbation theory
require the Gruneisen parameter to be zero along various directions in
reciprocal space. 

\begin{figure}
\begin{tikzpicture}
    \node at (0.0,6.17) { \resizebox{0.99\linewidth}{!}{\includegraphics{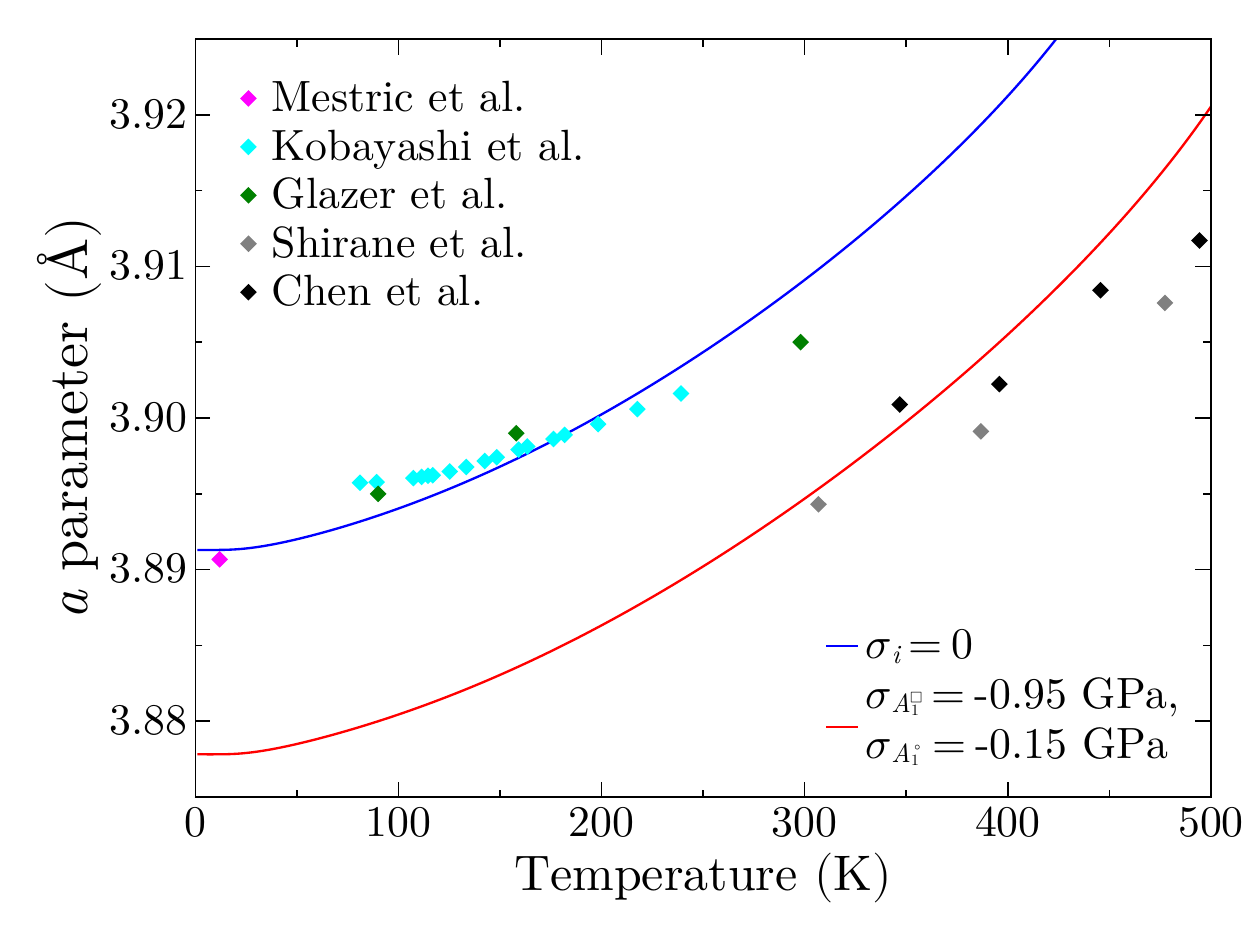}} };
    \node at (-3.8,9.0) {(a)};
    \node at (0,0) { \resizebox{0.99\linewidth}{!}{\includegraphics{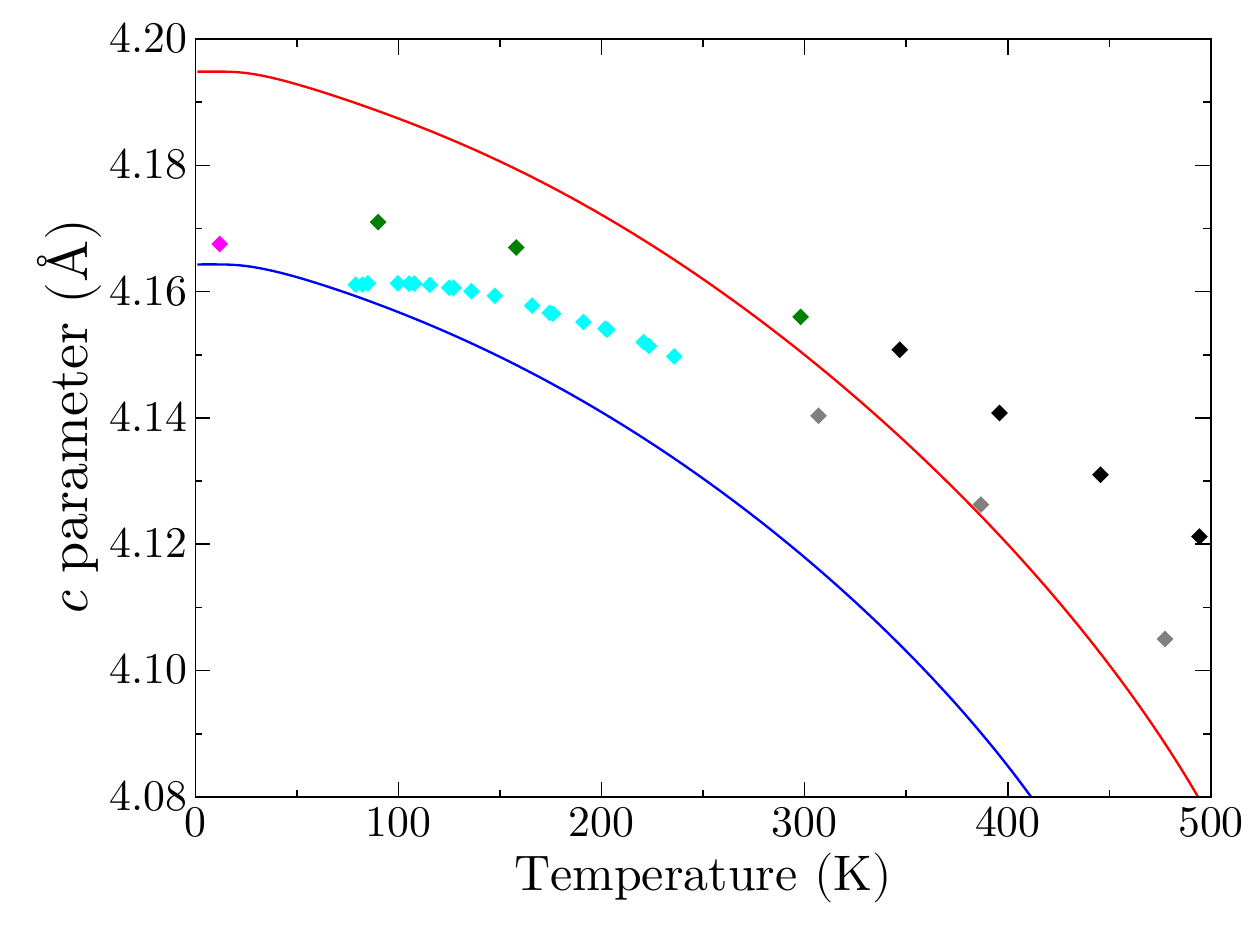}} };
    \node at (-3.8,2.9) {(b)};
\end{tikzpicture}
    \caption{
        The $a$ (panel $a$) and $c$ (panel $b$) lattice parameters computed with the QHA (lines) under unstressed (blue)
        and stressed (red) conditions compared with previous experimental measurements (diamonds)
        \cite{Shirane1951265,Glazer19781065,Kobayashi19833866,Mestric2005134109,Chen2005231915}.
    }
    \label{fig:thermalexp}
\end{figure}

Having computed the strain dependence of the phonons, we apply the generalized
quasiharmonic approximation to compute the $a$ and $c$
lattice parameters at finite temperature and stress (see Fig.
\ref{fig:thermalexp}). 
We compare with
experimental measurements at various temperatures under unstressed conditions 
\cite{Shirane1951265,Glazer19781065,Kobayashi19833866,Mestric2005134109,Chen2005231915}.
The $a$ and $c$ lattice parameters are in good agreement within the experimental measurements, 
as there is at most 0.5\% difference between various measurements at any given
temperature.
Computation of the lattice parameters at a given temperature $T$ and stress
$\sigall$ is achieved through the Biot strain function,
\begin{align}
\label{eq:strainmap}
\tilde{\epsall}(T,\sigall), \hspace{3mm} \textrm{where} \hspace{3mm} 
\tilde{\sigall}(T,\tilde{\epsall}(T,\sigall)) = \sigall.
\end{align}
where definitions and notation are equivalent to Ref.
\cite{Mathis2022014314} (see Eqs. 21-26).
The phonons are computed on a grid of strains and compared with a Taylor series
expansion of the phonons in strain, where convergence of the thermal expansion is
achieved when including first, second, and third strain derivatives within the Taylor series
(see Sec. I of the SM \cite{SM}).
The crystal structure predicted at $T=1$ K by the QHA differs from the values
obtained from DFT relaxations due to zero point motion, and are compared with 
experimental measurements at $T=12$ K \cite{Mestric2005134109} (see Table \ref{table:latticeparams}). 
The predicted shift in lattice parameters due to zero-point motion yields
remarkable agreement with the values obtained from experiment, however there
are discrepancies in the predicted direct atomic coordinates.
The change in lattice parameters with temperature is overestimated by the
quasiharmonic approximation which is well known for anharmonic materials
\cite{Allen2015064106,Masuki2022064112}.
Under the application of $\sigma_{\aone}=-0.95$ GPa and $\sigma_{\onea}=-0.15$
GPa, the $a$ and $c$ lattice parameter computations agree with experimental
measurements at room temperature. 

Having computed the lattice parameters, 
we now discuss the strain curvature of the free energy at finite temperature and stress.
There are three experimentally relevant quantities related to the
free energy curvature at finite stress \cite{Wallace19980486402126}: 
the free energy curvature $C_{ij}$, the elastic wave propagation coefficient
$S_{ij}$, and the stress-strain coefficient $B_{ij}$.
Additionally, elastic constants in ferroelectrics can be measured under boundary conditions of
constant electric field $E$ or constant electric displacement field $D$.
The relation between the two boundary conditions is given by \cite{Wu2005035105},
\begin{align}
    C_{ij}^{D}(T,\sigall) = & C_{ij}^E(T,\sigall) + \sum_{\alpha\beta}\big(e_{\alpha i}(T,\sigall) \times
    \nonumber \\
                            &  e_{\beta j}(T,\sigall) ((\hat{\epsilon}^{\infty,S}(T,\sigall))^{-1})_{\alpha\beta}\big),
\end{align}
where $\hat{\epsilon}^{\infty,S}$ is the relaxed-ion dielectric tensor at fixed
strain and $e_{\alpha i}$ is the relaxed-ion piezoelectric stress coefficient
\cite{deGironcoli19892853,Saghi-Szabo19984321,Saghi-Szabo199912771,Wu2005035105}.
Our evaluation of $\hat{\epsilon}^{\infty,S}$ and $e_{\alpha j}$ at finite
temperature $T$ and stress $\sigall$ is achieved by using the value at the
strain corresponding to $\tilde{\epsall}(T,\sigall)$.

\begin{figure}
\begin{tikzpicture}
    \node at (0.0,6.25) { \resizebox{0.99\linewidth}{!}{\includegraphics{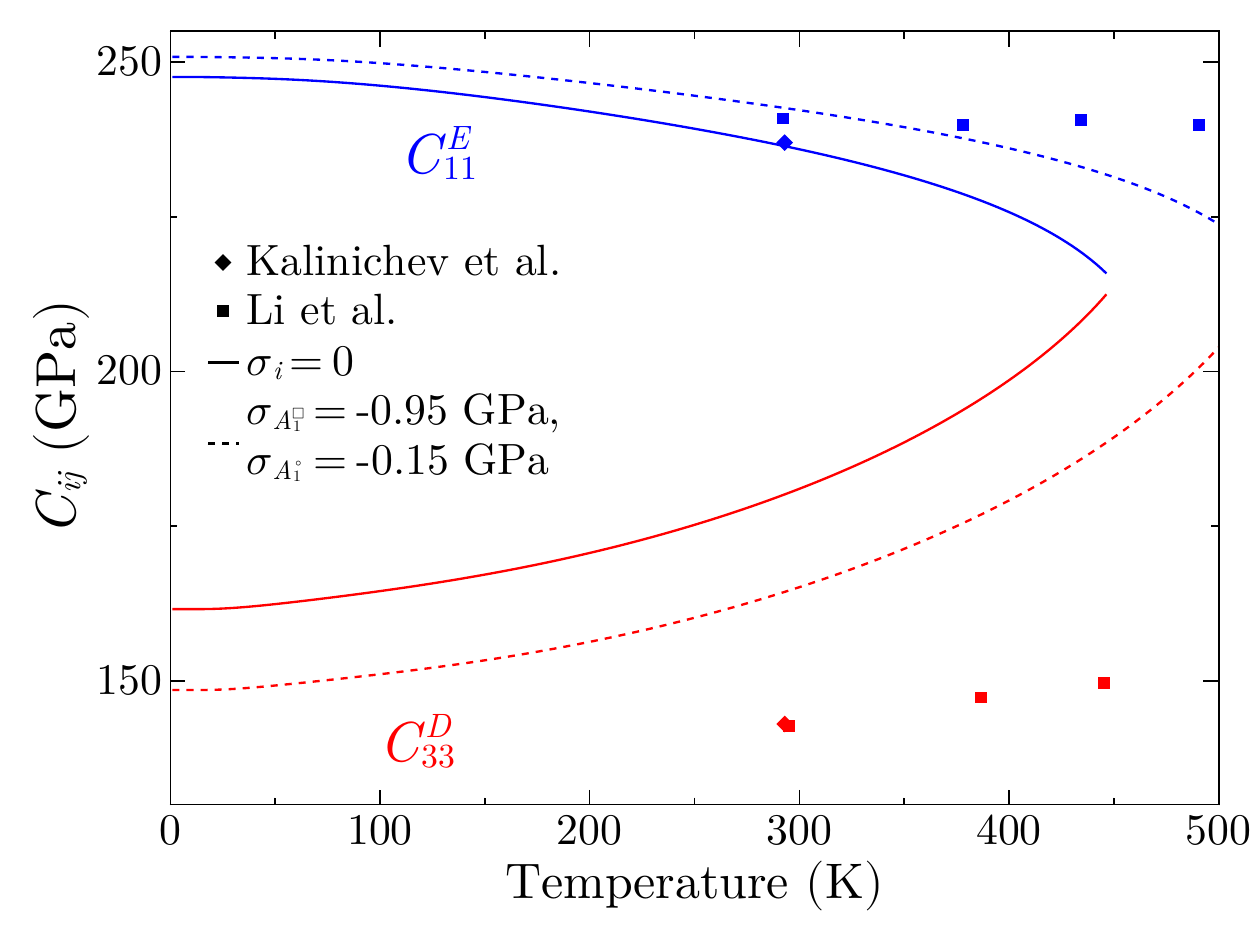}} };
    \node at (-3.9,9.05) {(a)};
    \node at (0,0) { \resizebox{0.99\linewidth}{!}{\includegraphics{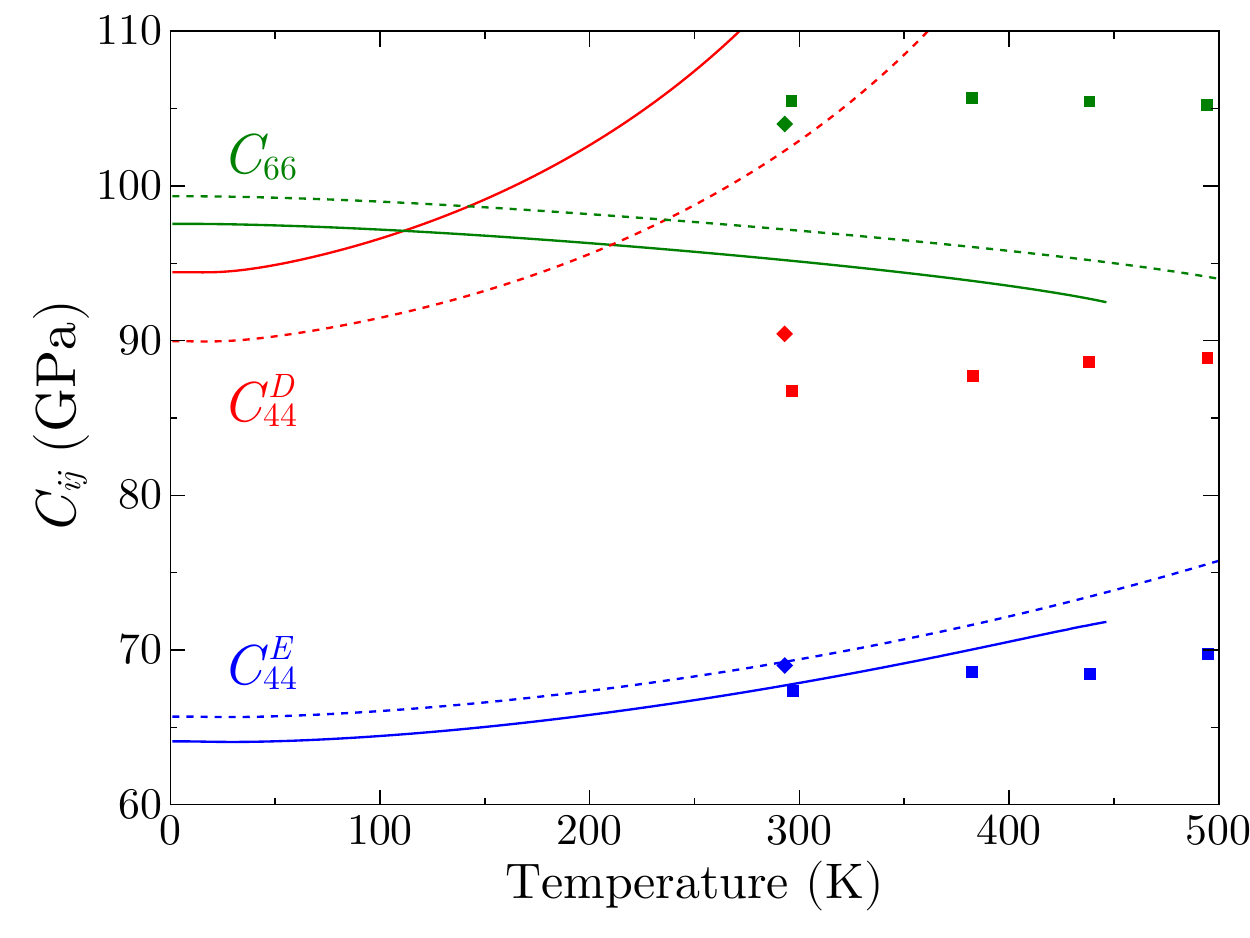}} };
    \node at (-3.9,2.9) {(b)};
\end{tikzpicture}
    \caption{
        Selected elastic constants under constant $E$ fields or $D$ fields
        computed with the QHA (lines) under unstressed (solid) and stressed
        (dashed) conditions compared with previous experimental measurements (markers)
        \cite{Kalinichev19972623,Li19961433}. The
        axial and shear elastic constants are shown in panels $a$ and $b$, respectively.
    }
    \label{fig:elastic}
\end{figure}

We compute the elastic
constant tensor at finite temperature and stress,
and compare with experimental values measured under unstressed conditions \cite{Li1993313,Kalinichev19972623,Li19961433}
(see Fig. \ref{fig:elastic}).
The elastic constants which were not measured at multiple temperatures are
shown in Sec. II of the SM \cite{SM}.
There is good agreement within the experimental values at room temperature, where
the largest disagreement is a 4\% difference in the measured $C_{44}^D$ values.
At room temperature, the QHA with PBEsol yields good agreement for $C_{11}^E$ and $C_{44}^E$,
however $C_{33}^D$ and $C_{44}^D$ are overestimated and $C_{66}$ is
underestimated. 
The change in the predicted elastic constants with temperature is greater than
the change observed in experiment, likely due to an overestimation of the
thermal expansion with temperature
(see Fig. \ref{fig:thermalexp}) and due to the neglect of explicit phonon
interactions within the QHA \cite{Allen2015064106,Masuki2022064112}.
The change in the constant $D$-field elastic constants with temperature is
significantly greater than the change seen in experimental measurements, 
and is due to the temperature dependence of $e_{ij}$ and $\hat{\eps}^{\infty,S}_{ij}$. 
These relaxed-ion quantities depend on the $\Gamma$-point dynamical matrix, and 
therefore the temperature dependence is overestimated due to the aforementioned
volume and anharmonic effects.

\begin{figure}
\begin{tikzpicture}
    \node at (0.07,9.90) { \resizebox{0.975\linewidth}{!}{\includegraphics{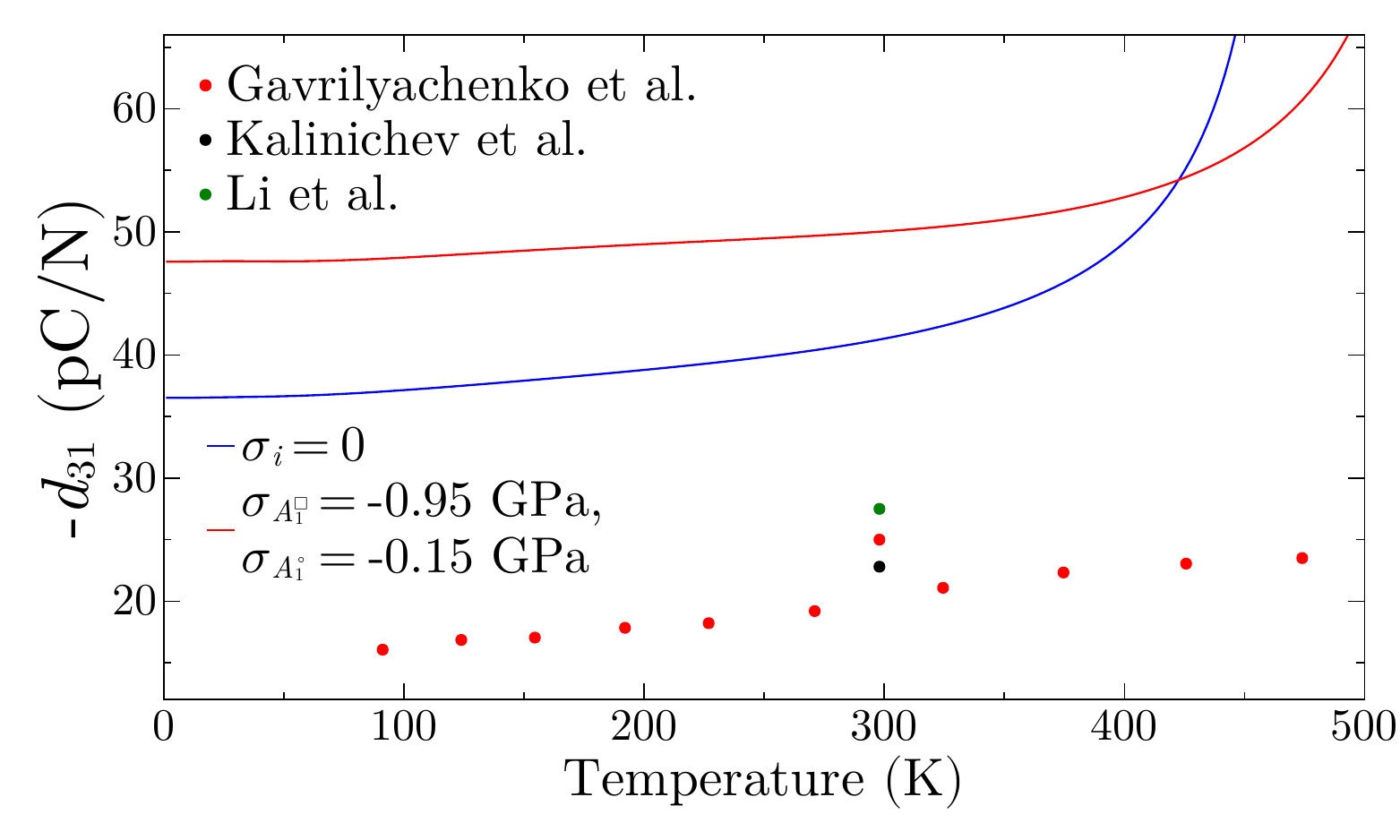}} };
    \node at (-3.9,12.10) {(a)};
    \node at (0.0,4.95) { \resizebox{0.99\linewidth}{!}{\includegraphics{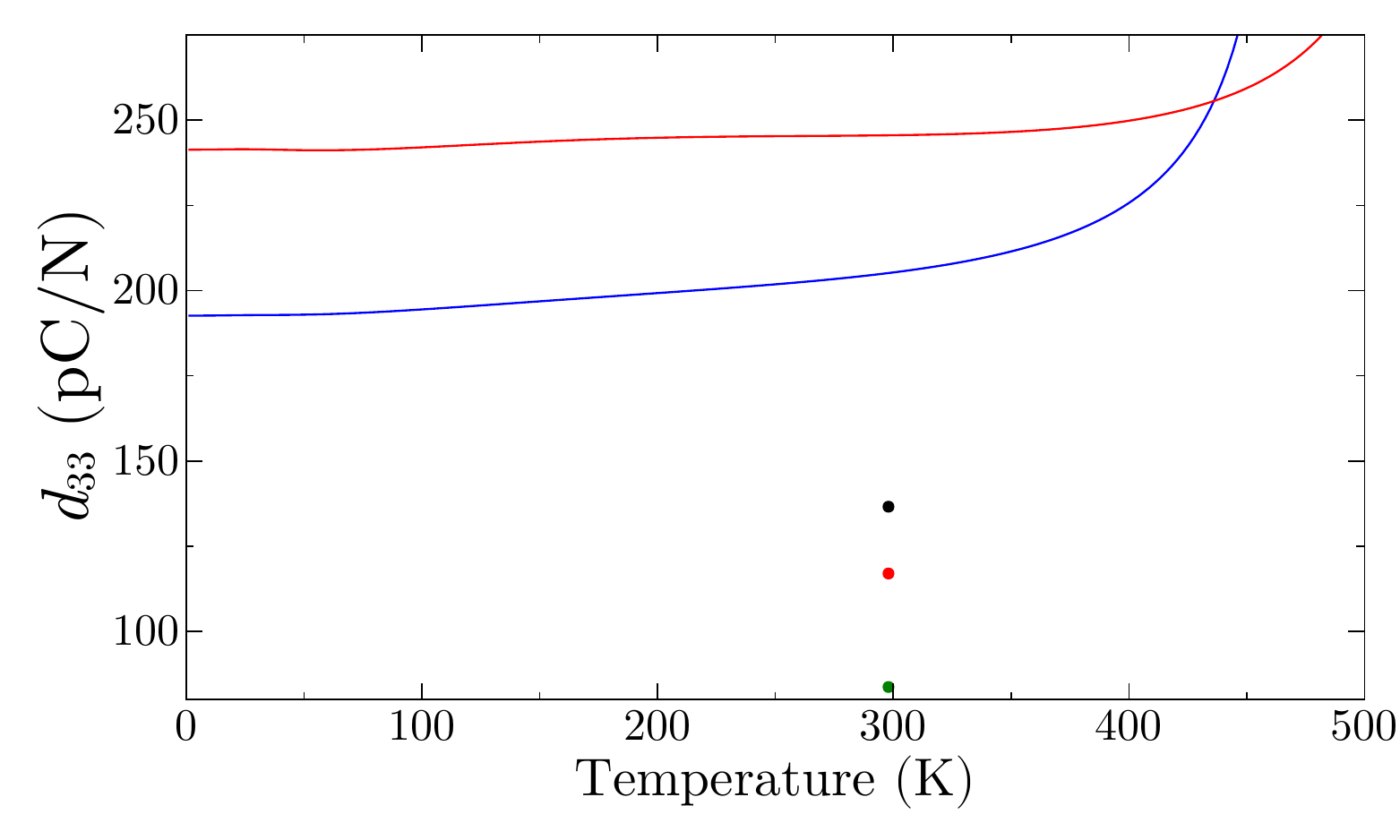}} };
    \node at (-3.9,7.10) {(b)};
    \node at (0,0) { \resizebox{0.99\linewidth}{!}{\includegraphics{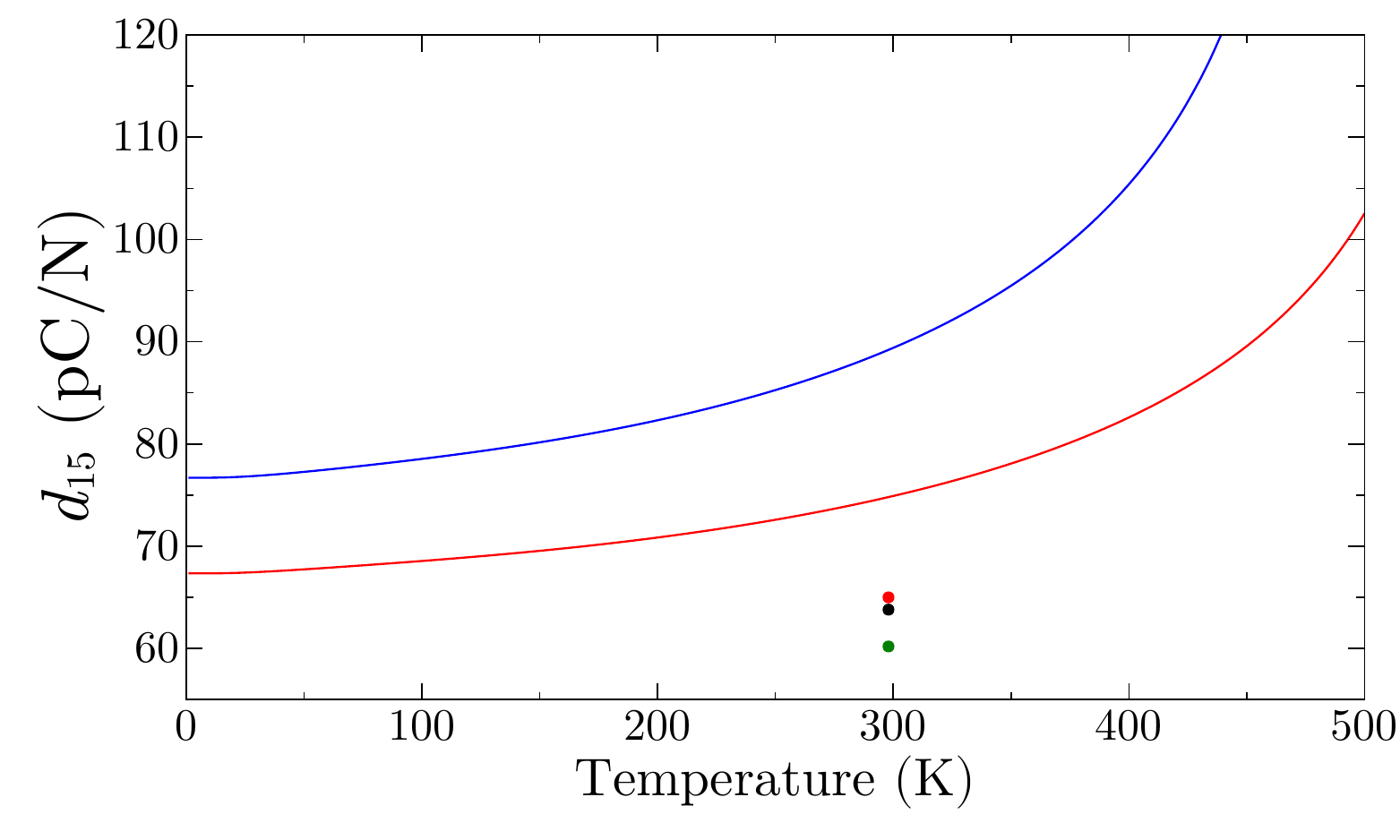}} };
    \node at (-3.9,2.20) {(c)};
\end{tikzpicture}
    \caption{
		Piezoelectric strain coefficients $-d_{31}$,  $d_{33}$, and
$d_{15}$ computed with QHA (lines) under unstressed (blue) and stressed (red)
conditions compared with existing experimental measurements (circles)
\cite{Li19961433,Kalinichev19972623,Gavrilyachenko19711971} are shown in panels
$a$, $b$, and $c$, respectively.
    }
    \label{fig:piezo}
\end{figure}
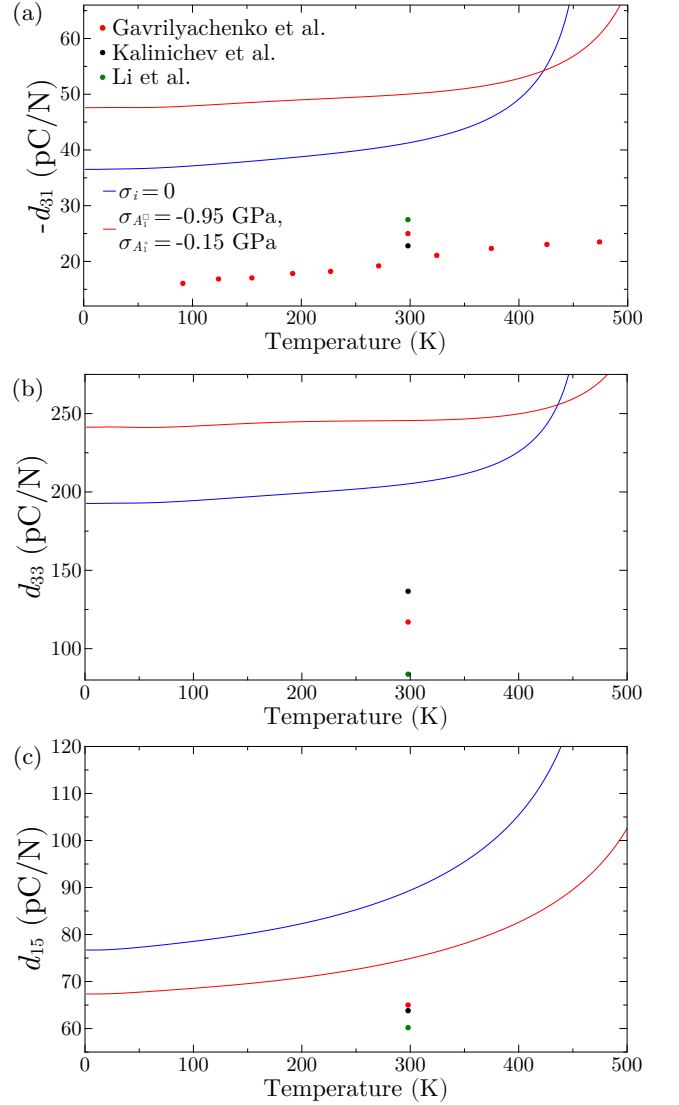

We proceed with the computation of the piezoelectric strain coefficients as a
function of temperature and stress, and compare with existing experimental
values measured under unstressed conditions 
\cite{Li19961433,Kalinichev19972623,Gavrilyachenko19711971}
(see Fig. \ref{fig:piezo}).
The piezoelectric strain coefficients $d_{ij}$ are constructed using the
elastic constant tensor and the piezoelectric stress coefficients
\cite{Nye1985,Wu2005035105},
\begin{align}
    d_{\alpha i}(T,\sigall) = \sum_j 
    ((\hat{B}^{E}(T,\sigall))^{-1})_{ij} 
    e_{\alpha j}(T,\sigall),
    \label{eq:piezo}
\end{align}
where $\hat{B}_{ij}^{E}$ denotes the stress-strain coefficient
under a constant electric field. 
The experimentally measured values of the piezoelectric strain coefficients 
\cite{Li19961433,Kalinichev19972623,Gavrilyachenko19711971}
show quantitative inconsistency, as the values vary as much as 50\% for
$d_{33}$ and 20\% for $d_{31}$.
Our quasiharmonic predictions of the $d_{33}$ and $d_{31}$ 
overestimate the experimental measurements, and the discrepancy can be explained by 
differences in the $C_{\aone\onea}^E$ and $C_{33}^E$ elastic constants. 
We compute values of $d_{33}=$ 201.6 pC/N and $d_{31}=-41.1$ pC/N using Eq. \ref{eq:piezo}, in good
agreement with our QHA predictions, by using the experimental values from Ref.
\cite{Kalinichev19972623} and substituting the values of $C_{\aone\onea}^E$ and
$C_{33}^E$ to the room temperature computed values of 108.3 GPa and 56.2
GPa, respectively.
The temperature dependence of the $d_{31}$ piezoelectric coefficient agrees with 
the only temperature dependent experimental measurement \cite{Gavrilyachenko19711971} 
from low temperatures up to room temperature, where the QHA predictions 
increase rapidly. 

In summary, we have demonstrated the application of the generalized
quasiharmonic approximation to a non-cubic crystal, ferroelectric PbTiO$_3$,
under conditions of finite temperature and stress.
The irreducible derivative approach to computing phonons from finite difference
yields the strain dependent phonons, where dipole-quadrupole effects are
successfully incorporated in the Fourier interpolation.
The thermal expansion, elastic constants, and piezoelectric strain coefficients
are computed at finite temperature and stress.
The temperature dependence of the thermal expansion and elastic constants at zero
stress are over estimated by the quasiharmonic approximation, illustrating the
need to solve the vibrational Hamiltonian of PbTiO$_3$ using a strain-dependent
theory which explicitly accounts for phonon interactions. 
Our observed limitations of the QHA are not unexpected, as
discrepancies of the QHA are well known in various anharmonic materials
\cite{Allen2015064106,Masuki2022064112}. 
Advances in the computation of finite temperature vibrational properties from
DFT using more sophisticated approximations than the QHA have been achieved
\cite{Souvatzis2008095901,Souvatzis2009888,Hellman2011180301,Errea2014064302,Tadano2015054301,Xiao2023094303}, 
however we are not aware of the application of any of these theories to the
computation of the elastic constant tensor at finite temperature.
These more advanced theories can be straightforwardly applied as a function of strain to
compute thermal expansion and elastic constants at finite temperature and stress using the general
formalism outlined previously \cite{Mathis2022014314}, which will be the subject of future work.

This work was supported by the grant DE-SC0016507 funded by the U.S. Department
of Energy, Office of Science.
This research used resources of the National Energy Research Scientific
Computing Center, a DOE Office of Science User Facility supported by the Office
of Science of the U.S. Department of Energy under Contract No. DE-AC02-05CH11231.

\bibliographystyle{apsrev4-2}
\bibliography{main}

\end{document}